# On the effectiveness of the thermoelectric energy filtering mechanism in low-dimensional superlattices and nano-composites

Mischa Thesberg[1,3], Hans Kosina[1], and Neophytos Neophytou[2]

[1]Institute for Microelectronics, TU Wien, Austria

[2]School of Engineering, University of Warwick, Coventry, CV4 7AL, UK

[3]e-mail: thesberg@iue.tuwien.ac.at

## Abstract

Electron energy filtering has been suggested as a promising way to improve the power factor and enhance the *ZT* figure of merit of thermoelectric materials. In this work we explore the effect that reduced dimensionality has on the success of the energy-filtering mechanism for power factor enhancement. We use the quantum mechanical non-equilibrium Green's function (NEGF) method for electron transport including electron-phonon scattering to explore 1D and 2D superlattice / nanocomposite systems. We find that, given identical material parameters, 1D channels utilize energy filtering more effectively than 2D as they: i) allow one to achieve maximal power factor for smaller well sizes / smaller grains (which is needed to maximize phonon scattering), ii) take better advantage of a lower thermal conductivity in the barrier/boundary materials compared to the well/grain materials in both: enhancing the Seebeck coefficient; and in producing a system which is robust against detrimental random deviations from optimal barrier design. In certain cases we find that the relative advantage can be as high as a factor of 3. We determine that energy-filtering is most effective when the average energy of carrier flow varies the most in the wells and the barriers along the channel, an event which appears when the energy of the carrier flow in the host material is low and when the energy relaxation mean-free-path of carriers is short. Although the ultimate reason these aspects, which cause a 1D system to see greater relative improvement than a 2D, is the 1D system's van Hove singularity in the density-of-states, the insights obtained are general and inform energy-filtering design beyond dimensional considerations.

# I. Introduction

The ability of a material to convert heat into electricity is quantified by the dimensionless figure of merit $ZT=\sigma S^2 T/\kappa$, where $\sigma$ is the electrical conductivity, $S$ is the Seebeck coefficient, and $\kappa$ is the thermal conductivity. Recently, large improvements in the $ZT$ of nanostructures were obtained through drastic reduction in thermal conductivity [1-3]. Similarly, efforts are underway to achieve power factor (PF) $\sigma S^2$ improvements and increase the $ZT$ even further, especially utilizing low-dimensional materials [4].

One of the strategies to improve the power factor, which has attracted significant attention, is energy-filtering in nano-composites and superlattices (SLs) [5-18]. The primary motivation for using such systems is that they are able to scatter long wavelength phonons on their many internal boundaries and greatly reduce the thermal conductivity [19-23]. However, due to the difference in scattering mean-free-paths of phonons and electrons, the expectation is that these boundaries/barriers will harm the electrical conductivity less. Furthermore, they have the potential to increase the Seebeck coefficient through the mechanism of energy-filtering via two mechanisms [5, 6, 24]: i) only carriers with high enough energies can overpass the barriers, and ii) when the barrier material has a lower thermal conductivity than the host bulk material, its higher $S$ is weighted more in determining the overall $S$, without loss in conductance. Thus, although the greatest benefit of such systems is their reduced thermal conductivity, they can also mitigate the power factor loss associated with the insertion of such scattering centers and in some cases may even improve it. Indeed, theoretical works by us and others indicate that energy filtering by a single potential barrier, or multiple barriers within an SL material system, can indeed provide power factor improvements, potentially up to 30% depending on what one compares against [25-29].

To-date, however, with the exceptions of Refs. [16, 30], only improvements in the Seebeck coefficient, but not the power factor, have been experimentally observed [7]. In general, power factor enhancement is only realizable if the conductance is not overly reduced by the addition of these barriers. Due to the interrelated nature of the Seebeck coefficient and conductivity, determining the ideal form of such structure geometries and barriers is non-trivial. Simulations have shown that the optimal design of a 1D SL geometry has peculiar features, dictating stringent requirements on 'effective' potential barrier heights and fine-tuning of the sizes of wells and barriers to correspond to the energy relaxation and tunneling probabilities of charge carriers [5, 16, 26, 27]. In Ref. [28] we showed that a possible reason for the general absence of power factor improvements could be unintended random variations in the heights of the barriers, away from the intended ideal, in the channel (as is common in practice), which were shown to be especially detrimental to the power factor. On the other hand, reasonable variations in barrier shape (deviations from perfect square barriers – as long as they are sufficiently thick to prevent tunnelling), as well as well and barrier sizes and position, do not cause significant power factor degradation [28].

The latter is important, because it suggests that studies on simplified SL geometries (in which barriers and wells are precisely placed), also provide insights into nano-composite structures (with 'on-average' placement) at first order. However, energy filtering channels can be built in 1D (superlattice naowires), 2D (superlattices) or 3D (nano-composites). Although there has long been arguments made for a beneficial effect of lower-dimensionality on the Seebeck coefficient in uniform structures [4], the effects of dimensionality in these filtering nano-structured systems with both spatially varying conduction bands and thermal conductivity, in addition to random variations in barrier shape, have not yet been discussed. In these energy-filtering structures, carrier behavior is far from equilibrium, with the average energy of the carrier flow rising and falling throughout the material as the carriers pass over potential barriers and then relax towards equilibrium in the wells. It is not yet clear if lower dimensionality also benefits the filtering mechanism as well, and quantifying this is the central focus of this paper.

Therefore, in the ideal energy-filtering system, other than the appropriate optimized geometrical features and barrier heights, we postulate the following criteria for an 'effective' energy-filtering strategy: i) feature sizes (i.e. barrier separation) should be as small as possible, to best scatter the dominant heat-carrying phonons and introduce larger heat resistance; ii) as a reduced thermal conductivity of the barrier material is an attractive means to achieve easy power factor enhancement, a system should receive maximal power factor enhancement with respect to this reduction, and iii) random and imperfect barrier heights should not have a substantial effect on the power factor. We omit the consideration of random barrier placement, shape and width as previous work [28] suggests in has only a small effect. Thus, the question we wish to address is: Will a lower-dimensional channel (1D), or a higher-dimensional channel (2D) utilize energy-filtering more effectively? I.e., will the power factor improve (or suffer less) once an SL is formed using a 1D channel or using a 2D channel, given the same set of material parameters? The goal is *not* to investigate in absolute terms if a 1D SL channel will provide higher power factors compared to a 2D SL channel, but rather if the additional effort of nano-structuring barriers in pays off more in 1D or 2D.

Owing to the highly non-equilibrium flow in such structures, here we employ quantum transport simulations based on the Non-Equilibrium-Green's Function (NEGF) method to isolate the effect of reduced dimensionality on the energy-filtering mechanism. We show that, for the same material parameters, a 1D channel sees greater power factor enhancement (or less degradation) for all three criteria mentioned above: i) for smaller SL periods; ii) from having a barrier material of lower thermal conductivity; and iii) in the face of random and imperfect barrier profiles that realistically occur in nano-composites. Although the physical reasons for the advantage of 1D ultimately stem from the shape of the 1D density-of-states, our conclusions provide important design insights with respect to the optimal design of the energy profile of the carrier flow along the SL or nano-composite channel, regardless of dimensionality.

## II. Approach

For the transport calculations, we use here the NEGF approach, including the effect of electron scattering with acoustic and optical phonons [31-32]. The system is treated within the effective mass approximation with uniform mass $m^* = m_0$, where $m_0$ is the rest mass of an electron. The effect of electron-phonon scattering is modeled by including a self-energy on the diagonal elements of the Hamiltonian [33]. For simplicity we choose the strength of the electron-phonon coupling for both acoustic and optical phonons to be the same, $D_0 = 0.0016$ eV$^2$ [31]. The optical phonon energy we consider is 60 meV. Throughout this work a Fermi-level of $E_F = 0.075$ eV was used, being a value which is $3k_B T$ above the conduction band, and thus metallic, but close enough to it to be a plausible model of a highly degenerate semiconductor.

The parameters used do not refer to a specific material, but are intended to have a broader applicability, as the primary goal is to isolate the effect of dimensionality alone. Any qualitative conclusions drawn should be independent of any specific degenerate semiconductor material and instead only reflect dimensional effects as it relates to energy-filtering. That said, however, we have chosen parameters that somehow reflect the transport features of usual semiconductor materials. The strength of the phonon scattering is chosen to have the same strength for acoustic and optical phonons for simplicity, but this assumption is not dissimilar from semiconductors like silicon, where their relative strengths are of the same order. The amplitude of the strength is chosen such that the mean-free-path for scattering is ~20 nm, which is also a usual case for common semiconductors; and the optical phonon energy of 60 meV chosen is actually the same as the prominent optical phonon energy in Silicon. Thus, even qualitatively, we have chosen our parameters to reflect usual materials employed for nanostructured TEs.

The power factor, $GS^2$, was obtained from the expression:

$$I = G\Delta V + SG\Delta T. \qquad (1)$$

For each value of the power factor, the calculation was run twice, initially with a small potential difference and no temperature difference ($\Delta T=0$), which yields the conductance ($G=I_{(\Delta T=0)}/\Delta V$), then again with a small temperature difference and no potential difference ($\Delta V=0$), which yields the Seebeck coefficient ($S=I_{(\Delta V=0)}/G\Delta T$). This method is validated in Ref. [25] and more details can be found in our previous work [29]. For computational reasons, the '2D channel', we use here has a width of $W$=12.5 nm. Figures 1a and 1b show the DOS versus energy in the 1D and the 2D channels we simulate, indicating that the DOS of the `2D channel' indeed possesses the almost constant DOS of a true 2D system. Figure 1c also shows a calculation of the thermoelectric power factor divided by width (and conductance divided by width and Seebeck in the inset) as a function of the channels' width, indicating saturation at widths wider than ~5nm. This saturation reflects the fact that as width is increased, the energy of 1D sub-bands lowers and their spacing in energy decreases. Thus, the number of 1D sub-bands in the energy window of interest increases (~12 can be seen in Figure 1b) such that the total DOS changes from that of 1D, with the shape of the inverse square root of energy, to a 2D constant DOS. The lack of width dependence of the power factor (divided by width) for large widths reflects the fact that the system has reached 2D rather than quasi-1D behaviour.

## III. Results

**Relative Filtering Benefits in 1D vs. 2D:** To compare the effectiveness of filtering in 1D and 2D, we perform the following: We consider a channel of constant length $L$=300nm with its conduction band at $E_C$ = 0 eV, and $E_F$ = 0.075 eV into the bands. We then place a single barrier in the middle of the channel of width $L_B$ = 42 nm. This width is chosen as it is larger than required for relaxation on the top of the barrier and can be evenly divided into 6 barriers of 7nm width (for our next step). Thus, it allows the comparison of two systems; one with a single barrier and one with many, with an identical amount of barrier material. Thus, the effect of the barrier spacing, and thus the effect of semi-relaxation, alone can be isolated by comparing the two. In practice, such barriers can be formed by alternation of materials within a superlattice, nano-compositing, by alloying the host material, electrostatically by doping variation or selective gating of specific regions, etc. The former are common techniques employed mainly to reduce the thermal conductivity of the materials and through this improve the *ZT* figure of merit. We then raise the single potential barrier gradually to achieve energy filtering and at every instance we compute the thermoelectric coefficients: electrical conductance *G*, Seebeck coefficient *S*, and power factor $GS^2$ for both 1D and 2D channels. These are plotted in Fig. 2a for 1D and Fig. 2b for 2D channels versus the barrier height $V_B$. The insets show the *G* and *S*. The dashed-dotted lines indicate the PF of the reference channel with $V_B$ = 0 eV. Two important observations can be made from this comparison: i) The relative maximum increase in the PF in 1D is just slightly higher than 2D (22.5% vs. 21%), but, more importantly, ii) the PF is optimized at different barrier heights for the 1D (~10 meV *above* $E_F$, inset of Fig. 2a) and the 2D (~10 meV *below* the $E_F$, inset of Fig. 2b) cases. The van-Hove singularity of the 1D bands lowers the average energy of the current flow, which requires a higher $V_B$ to reach optimal PF (and introducing a ~37% increase in $S$ in 1D versus ~31% in 2D, but a ~35% drop in $G$ in 1D versus 29% in 2D). In 2D the current flow is naturally higher in energy because of the more-or-less constant DOS. Although the relative PF difference between 22.5% and 21% seems small, this originates from filtering in a small part of the channel only. In fact if we triple the barrier size to $L_B$=126nm (almost half of the entire channel), these numbers change to 54% for 1D and 43% for 2D, which indicates that the higher rise in energy flow provides 1D a clear relative PF advantage in utilizing energy filtering (of 11% points with respect to the original channel PF).

After examining a single barrier, which under optimal $V_B$ provides some small relative advantage in 1D over 2D, we move to examine filtering in superlattices. For this, we split the thick barrier into *six* smaller barriers and begin to increase the separation between them ($L_W$), creating wells. This is shown schematically in Fig. 3 from a top view as well as in the inset of Fig. 4a in cross-section. We then compute the TE coefficients as $L_W$ increases from $L_W$ = 10 nm to 50 nm, forming a superlattice whose 'barrier material amount' stays the same at 42 nm, and examine the impact on the power factor as it is distributed in different ways in a channel of fixed length. The barrier size of $L_B$ = 7 nm is thick enough to prevent quantum tunneling that degrades the power factor severely [27-28]. We calculate the PF for both the 1D and 2D channels and compare its relative improvement in these SL geometries to the channel with a single thick barrier in Fig. 2. The reason we do this is to isolate the effect of relaxation in the wells from the effect of average energy enhancement. In reality the optimal $V_B$ for maximum PF in these 1D and 2D SL geometries turned out to be $V_B = E_F + k_B T$ and $V_B = E_F$, respectively. Thus, in the rest of the paper we use these values that favor the many barrier case (rather than the single barrier case).

The simulation results for the *relative* change in the $G$, $S$, and PF versus the well length $L_W$ are shown in Fig. 4a-c, respectively (we divide by the corresponding PF of the single barrier geometry). The solid lines (squares) show the results for the 1D channel whereas the dashed lines (diamonds) for the 2D channel. It is important to clarify here that what is shown in Fig. 4 is *not* a comparison of the absolute value of thermoelectric parameters in a 1D channel versus those in 2D. Such a comparison would not even be meaningful in the present context as the conductance of a 2D material scales with width and can thus be made arbitrarily large. Rather what is shown is a comparison of a 1D channel with a regular periodic array of potential barriers, to a 1D channel with the same amount of foreign material placed in the center of the channel (and the same for 2D). And thus, as the comparison is between a multi-barrier energy-filtering channel versus a single-barrier energy-filtering channel of the same dimensionality, the resulting ratio is unitless and can be compared between dimensions. The intent here is not to address whether a 2D superlattice is superior to a 1D superlattice, but rather whether energy-filtering as a design strategy is more effective in 2D versus 1D. In other words, what this figure describes is as follows: "Given that you already have a 2D or 1D system, what enhancement in the Power Factor can you expect from energy-filtering in a SL geometry?"

The relative change in the electrical conductance (Fig. 4a) shows that as the barriers are spread in the channel and more of the channel area is occupied by barriers, $G$ drops, as expected. On the other hand, the Seebeck coefficient in Fig. 4b follows the reverse trend and is increased as the barriers are more spread in the channel because this increases energy filtering over a larger length, again as expected.

The relative power factor (PF) changes, however, in Fig. 4c reveal some interesting features. A clear improvement is observed as $L_W$ is increased and the barriers are spread in the channel forming the superlattice. In the 1D case, however, for the same geometries ($L_W$) the relative improvements are more than in 2D (ratios up to ~3× in some instances), with the 2D only reaching the 1D improvements at maximum $L_W$. This is a clear indication that energy filtering is favored in 1D SLs for smaller well sizes ($L_W$ < 50 nm), compared to 2D SLs for which the benefits of energy filtering are maximized at much larger well sizes, for the same set of material parameters. SLs with smaller well regions would potentially favor thermoelectric materials, as they also provide larger heat resistance and smaller thermal conductivities. Looking at Fig. 4a and 4b, it is clear that the advantage of the 1D SL in utilizing energy filtering, resides in the fact that in 1D, the electrical conductance suffers less than in 2D with the introduction and spread of the barriers. The relative improvement of the Seebeck coefficient is actually higher in 2D. Overall, however, the PF in 1D is improved more, indicating that the changes that are introduced in G dominate the behavior of the PF.

It is quite interesting to mention here that no matter if we use a single barrier, or a SL geometry, the 1D channels utilize the filtering mechanism more effectively. For example, the relative PF advantage of 1D versus 2D in the case of a single large barrier of $L_B$ = 126 nm was 53% - 42% = 11% units compared to the reference empty channel as mentioned above. For the SL channel that extends a similar distance (~half of the entire channel), i.e. the case of six barriers of $L_W$ = 20 nm in Fig. 4c, the relative PF improvement in 1D is ~26%, more than double compared to the 12% for 2D, a difference of ~14% units. The relative advantage of 1D over 2D, either in a single barrier structure, or a SL structure is, thus, also of very similar value (11% higher and 14% higher).

## IV. Discussion

In Fig. 5a and Fig. 5b we explain the behavior of $G$ and $S$ by plotting the energy of the current flow along the channels' length in 1D and 2D, respectively, for the channel with six barriers. The blue lines indicate the *average* energy of the current flow. Two things are clear: i) *in the wells* there is significantly more energy relaxation in 1D compared to 2D (the energy relaxation length is extracted to be ~ 8 nm in 1D but ~13.5 nm in 2D, which is expected as the 1D DOS singularity provides more states for down-scattering), and ii) *in the barriers* the average energy of the current flow under optimal PF conditions is similar in 1D and 2D (slightly higher in 1D by about ~6 meV). This is clearly indicated in Fig. 5c, which combines the blue lines from Fig. 5a and Fig. 5b. Thus, $G$ suffers less in 1D (as shown in Fig. 4a) because the average energy is lower in the wells, closer to well equilibrium, but $S$ increases more in 2D (as shown in Fig. 4b) because the energy of the current flow is higher in the wells ($S$ is determined by the carrier energy as $\sim <E-E_F>$, a point discussed in the Appendix). Overall, however, the PF improvement is determined by $G$, and it is higher in 1D. For comparison, the green line in Fig. 5c shows the average energy of the current flow in the 1D channel with the same $E_F$ and $V_B$ as for the optimal 2D channel (i.e. $V_B = E_F$). The flow is lower than in 2D, which is why higher barriers are required in 1D to lift it up.

A more general observation at this point, is that filtering is more effective in channels in which the energy of the current flow is: i) closer to the band edge, and ii) has shorter energy relaxation length (as the 1D case). The introduction of barriers has 'more room' to raise the flow energy further and improve the *S*, which manifests as a modest relative increase in PF. Furthermore, within a SL geometry, the sharper energy relaxation in the wells can compensate for the loss in *G* due to the barriers by facilitating a more rapid return to equilibrium. Superlattice channels in which the current flow is further from the band edge and the energy relaxation lengths are longer, have reduced possibilities in utilizing filtering to improve the PF (as the 2D case), mainly because *G* suffers more compared to the benefits in *S*. Note that the conclusion we reach can be generalized to suggest that the benefits of filtering can be observed more in materials in which the current flows closer to the conduction band (as in 1D), rather than in higher energies (as in 2D and presumably 3D). In practice, purely 1D channels can be very difficult to achieve, but some of the light mass materials, i.e. III-Vs, InAs, InSb, BiTe, etc, could have nanowires built out of them, in which transport is dominated by a single 1D band even for channels with diameters up to several nanometers [34,35]. The 1D versus 2D relative comparison clearly shows the benefits of filtering in 1D is a result of a larger current flow energy variation (38 meV in 1D versus 17 meV in 2D as indicated in Fig. 5c). However, this is just one study case. The main argument can be extended to suggest that filtering in 2D materials such as quantum wells can provide relatively more benefits compared to filtering in 3D materials, where the current flow happens at higher energies. Our conclusion also suggests that materials with larger optical phonon scattering energies which allow more carrier down-scattering could also be more effective in energy filtering once nano-composites are built out of them. It is also important to stress though, that the relative filtering benefits observed in these SLs originate from how the shape of the 1D density-of-states energy function dictates the energy of the current flow, and not from how the sharp edge influences the Seebeck coefficient, as what proposed by Hicks and Dresselhaus but for uniform low-dimensional channels [4]. As a matter of fact, the main reason behind these relative benefits is the fact that in the potential wells, away from the filtering barrier regions, the lower energy of the current flow in 1D allows for improving the *conductance*, compensating by the reduction caused by the barriers.

**<u>Non-uniform thermal conductivity improves 1D filtering even more:</u>** Next, we explore another aspect of potential PF improvements in a SL geometry, which originates from the fact that $S$ can be further increased when the local thermal conductivity, $\kappa$, differs between the barrier and well regions. As discussed on various occasions, the overall $S$ is determined by integrating the local $S(x)$, weighted by the inverse of the local $\kappa(x)$ (or the temperature gradient, $dT/dx$) along the channel. Thus, regions of lower $\kappa$ have more weight in determining the overall $S$ [16, 25, 26]. This is because the Seebeck voltage drops more in regions of larger $dT/dx$, which are regions of lower $\kappa$ (see Appendix for a detailed derivation).

Figure 6 shows the *relative* improvement of the power factor in the SL channel for different barrier heights with different thermal conductivities in the barrier ($\kappa_B$) and the well ($\kappa_W$), versus the ratio of those thermal conductivities, $\kappa_B/\kappa_W$. Again, as earlier in Fig. 4, the goal is to examine what extent does the variation of $\kappa$ in the channel pays off in 1D versus 2D, and not to compare absolute power factor values. The SL channel simulated is the one with six barriers with wells of $L_W = 50$ nm at $E_F = 0.075$ eV, which had the largest relative improvements in Fig. 4. We keep the thermal conductivity of the barrier smaller than that of the well region (i.e. $\kappa_B/\kappa_W < 1$) in order to weigh more the superior $S$ of the barriers. The results for 1D are shown by the solid lines and for 2D by the dashed lines. Cases for different barrier heights are shown: $V_B = 0.05$ eV (blue lines), $V_B = 0.075$ eV (black lines - optimal case for 2D), $V_B = 0.1$ eV (red lines-optimal case for 1D), and $V_B = 0.125$ eV (green line - $k_BT$ above the optimal case for 1D). It is interesting to observe that even in this case, the 1D channel utilizes the difference between $\kappa_B$ and $\kappa_W$ more effectively, being able to provide ~50% more PF improvement for small $\kappa_B/\kappa_W$ ratios. The inset of Fig. 6 shows the same data but normalized to the $\kappa_B/\kappa_W = 1$ data point for each $V_B$ case. It clearly demonstrates that irrespective of $V_B$, the 1D channel utilizes a smaller $\kappa_B$ more effectively compared to 2D (all solid lines are higher than the dashed).

The reason for this is that in 1D the energy of the current flow is lower in the wells compared to 2D (and slightly higher in the barriers), and because of this, to begin with, in 1D $S$ is mostly determined by the barriers, whereas in 2D the wells also contribute substantially. Thus, since a smaller $\kappa_B/\kappa_W$ ratio weights $S$ in the barriers even more, the 1D channel is benefited more. This is also observed in Fig. 4b, which shows that the improvement in $S$ in 1D saturates earlier (becomes independent of $L_W$) compared to 2D. This is precisely because as $L_W$ is increased and the current energy relaxes lower in the 1D wells, those wells contribute less to $S$. Another interesting observation from the inset of Fig. 6, is that the higher the barrier, the larger the relative PF improvement, even for barriers higher than the optimal ones (i.e. the green-solid line is higher than the red-solid line).

The fact that the 1D channels, with the larger variation in the energy of the current flow along the transport direction, utilize filtering better by having their Seebeck coefficient determined mostly by the barriers, could lead to another important advantage in the design of SLs or nano-composite TEs. This is the relative immunity to unwanted barrier height variations. In a previous work we considered TE transport in SLs with uniform $\kappa$ along the SL, but considered reasonable values of variation in the barrier height, $V_B$. We showed that the power factor was drastically degraded, controlled mostly by the reduction in the conductance imposed by the highest barrier. Thus, we suggested that if one considers such a system, which includes variations in the barrier heights away from the optimal, then it is better to have non-optimal *lower* barriers than anomalously high ones, to avoid excessive reduction in conductance [28]. However, looking at Fig. 6 it can be seen that in both 1D and 2D systems, when the effect of reduced $\kappa_B$ is considered, the negative effect of a higher $V_B$ is mitigated, as the Seebeck coefficient will be additionally weighted by the lower $\kappa_B$ and ultimately that improvement compensates for the conductance loss (i.e. the green-solid line overpasses the black-solid line as the ratio is reduced, approaching the red-dashed line).

In Fig. 7 we performed such calculations, where we allow a rather large 30% statistical variations in the barrier heights, $V_B$, of the SLs of Fig. 6 for both 1D and 2D channels. The inset of Fig. 7 shows a few overlapping schematics of the barrier variations. In these cases we simulate 20 different channels for 1D and 2D and compute the PF in each case. Figure 7 shows the relative change in the thermoelectric power factor between a SL which has different thermal conductivity in the barriers ($\kappa_B$) and wells ($\kappa_W$), normalized to a superlattice with uniform thermal conductivity in all regions $\kappa_B/\kappa_W = 1$, plotted versus the ratio of the thermal conductivities in the barriers and wells (as in Fig. 6). The blue-dashed lines are results for 1D channels (with optimal $V_B = 0.1$ eV and no $V_B$ variations), whereas the red-dashed lines for 2D channels (with optimal $V_B = 0.075$ eV and no $V_B$ variations). The solid lines indicate the power factor of these two channels upon 30% variation in the barrier heights $V_B$ (blue-solid for 1D, and red-solid for 2D). First we consider the right side of this figure in the case of uniform thermal conductivity in the SL, $\kappa_B/\kappa_W = 1$. As expected, variations in $V_B$ degrade the power factor, which remains lower compared to the structures with uniform ideal $V_B$ for any $\kappa_B/\kappa_W$ ratio. The 1D channel is hurt more at $\kappa_B/\kappa_W = 1$ (blue-solid line is below the red-solid line) because the optimal barriers are higher to begin with anyway. As the ratio of the thermal conductivity is reduced, however (moving to the left of the graph), the 1D channel is able to compensate for that larger loss at larger $\kappa_B/\kappa_W$ ratios, and overpasses the 2D channel as the variation in thermal conductivities benefits 1D more than 2D. At the very left of the graph, for low $\kappa_B/\kappa_W$ ratios, both the 1D and 2D structures restore ~50% of the degradation that the $V_B$ variation causes (the solid lines approach at a large degree the dashed lines). The 1D however, sees this restoration at larger $\kappa_B/\kappa_W$ rations compared to the 2D channel, and the relative advantage is ~4× higher.

## V. Conclusions

In conclusion, we have investigated the effectiveness of the energy-filtering mechanism in improving the thermoelectric power factor in 1D versus 2D superlattices using quantum transport simulations. Ultimately, the question we addressed was the following: If one implements an energy filtering based TE material, does it pay off more to use a low-D or a higher-D material? We showed that, when compared at the same Fermi level and conduction band edge, 1D materials benefit more from energy-filtering because the presence of the van Hove singularity in their density-of-states energy function provides an overall lower average energy of the current flow, and shorter carrier relaxation lengths, compared to 2D materials. Thus, the introduction of energy-filtering barriers has more 'room' to raise the energy flow and improve the Seebeck coefficient, whereas the sharper energy relaxation in the wells allows for the conductance to still remain high, offering overall larger relative power factor improvements, compared to 2D materials. 1D superlattices or nano-composites allows for filtering to be more effective because: i) For optimal conditions they require shorter superlattice periods, or smaller average grain size in nano-composites (which is also beneficial in reducing thermal conductivity); ii) They utilize better the additional improvements in the Seebeck coefficient when the thermal conductivity of the barriers is smaller compared to that of the wells (which is a common case), and; iii) For these (usual case) materials where the thermal conductivity of the barriers is smaller compared to that of the wells, 1D provides larger immunity to the detrimental variations in the barrier heights, which could naturally appear. These are all features favorable for effective filtering, and provide larger relative power factor gains in 1D than in 2D. In general, we explained how this better utilization of the energy-filtering mechanism can be thought to originate from the larger variations of the average energy of the current flow as it travels through barriers and wells in 1D compared to those in 2D (by almost 2× larger).

As an example of how these insights could be applied to specific material systems, the InGaAs/InGaAlAs, or InAs/InGaAs (well-barrier) systems are good candidates. The barriers in both cases have a lower thermal conductivity compared to the wells, lower by a factor of or 5[36] in some cases. In addition, low-dimensional effects appear in these channels at 10s of nanometers,[37] due to their light effective mass, which make it technologically feasible to fabricate arrays of nanowires based on their superlattices [38]. On the other hand, the SiGe/Si (well-barrier) system, in which the thermal conductivity of the well is lower compared to the barrier, and in which low-dimensional effects appear at length scales below 10nm, making it technologically more challenging to reach, might have greater difficulty in taking advantage of the effects we describe here.

In general, these observation could potentially provide helpful generic guidance in picking better energy-filtering materials to create nano-composites out of, regardless of dimensionality.

*Acknowledgement:* MT has been partially supported by the Austrian Research Promotion Agency (FFG) project 850743 QTSMoS and partially by the Austrian Science Fund (FWF) contract P25368-N30. Some of the computational results presented have been achieved in part using the Vienna Scientific Cluster (VSC). NN has received funding from the European Research Council (ERC) under the European Union's Horizon 2020 research and innovation programme (grant agreement No 678763).

## Appendix A:

These results were obtained by assuming the lattice temperature ($T_L$) varies according to a simple thermal circuit model, as was discussed in [4]. In such a circuit of $n_B$ barriers of width $L_B$, where we assume that the temperature $T$, computed as $\Delta T = \int_0^L (dT_L / dx) dx$, is simply composed of regions of two different temperature drops per distance: $dT_L/dx|_{\kappa_B}$ (in the barriers) and $dT_L/dx|_{\kappa_W}$ (in the wells). The entire temperature drop across the channel is decomposed as:

$$\Delta T = \frac{dT_L}{dx}\bigg|_{\kappa_B} nL_B + \frac{dT_L}{dx}\bigg|_{\kappa_W} (L_{ch} - nL_B) \qquad (A1)$$

At an interface between different materials, heat flux is conserved and we have:

$$-\kappa_B \frac{dT_L}{dx}\bigg|_B = -\kappa_W \frac{dT_L}{dx}\bigg|_W \qquad (A2)$$

From (1-2), one arrives at the expression:

$$\frac{dT_L}{dx}\bigg|_W = \frac{\Delta T}{L - n_B L_B \left(1 - \frac{\kappa_W}{\kappa_B}\right)} \qquad (A3)$$

and a similar expression for $\dfrac{dT_L}{dx}\bigg|_B$.

From knowledge of the total temperature difference and the temperature gradients in the two regions one can determine $T_L(x)$ across the whole channel. The difference in thermal conductivities then affects the relative steepness of $T_L(x)$ in the barrier region versus the wells. We then note that the Seebeck coefficient can be represented as:

$$S = \frac{\int_0^L S(x)(dT_L/dx)\,dx}{\Delta T} = \frac{\int_0^L \left[\langle E(x) - E_F \rangle / qT_{L(x)}\right](dT_L/dx)\,dx}{\Delta T} \qquad (A4)$$

where

$$\langle E(x)\rangle = \int_E EG(E)\left[f_S(E)-f_D(E)\right]dE \,/\, \int_E G(E)\left[f_S(E)-f_D(E)\right]dE \quad \text{(A5)}$$

This is thus just the regular expression for the Seebeck coefficient. Thus, by using the spatially and energy resolved current information obtained from an NEGF simulation, it is possible to calculate $\langle E(x)\rangle$ and then S(x), and therefore the Seebeck coefficient of the whole system, *S*, by summing up every spatial point in the transport direction. This means of determining the Seebeck through a summation of the average energy is an alternate means from the "two-runs'" method we used in the bulk of this work and was found to be in strong agreement.

Thus, we see that the action of a lower thermal conductivity is that it ultimately increases the contribution of the barrier region where $\langle E(x)-E_F\rangle$ is highest and thus improves the overall Seebeck without having any first order effect on the conductivity.

Figure 1:

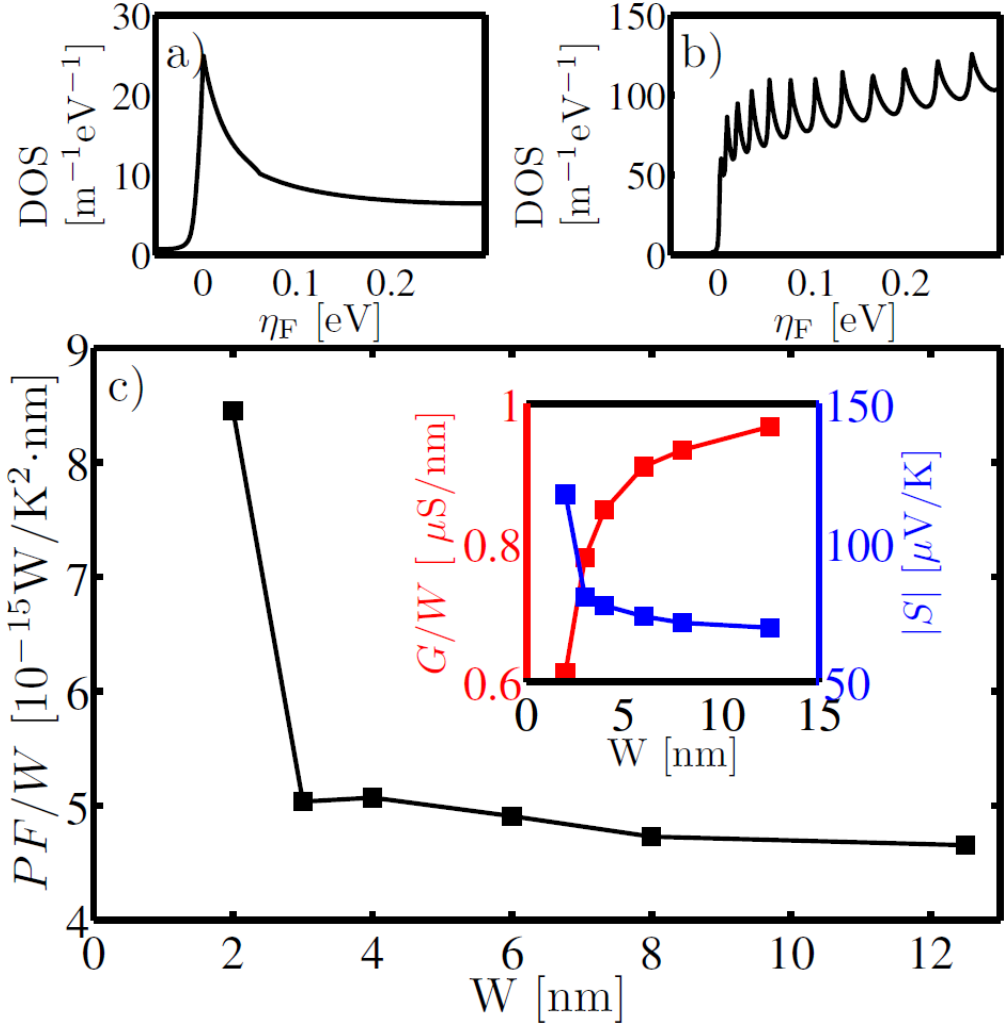

Figure 1 caption:

(a)The density-of-states (DOS) versus energy in the 1D channel we simulate. (b) The DOS in the 12.5 nm wide channel we simulate, which resembles 2D, and referred to as '2D channel' in the paper. (c) The thermoelectric power factor divided by width of the channel as the width increases from $W$ = 2.5 nm to $W$ = 12.5 nm. The Fermi level is at $E_F$ = 0.075 eV. Inset: The electrical conductance divided by width and Seebeck coefficient. All thermoelectric coefficients saturate after $W$~5nm within a few percentage points, which justifies our use of $W$ = 12.5 nm as '2D'.

Figure 2:

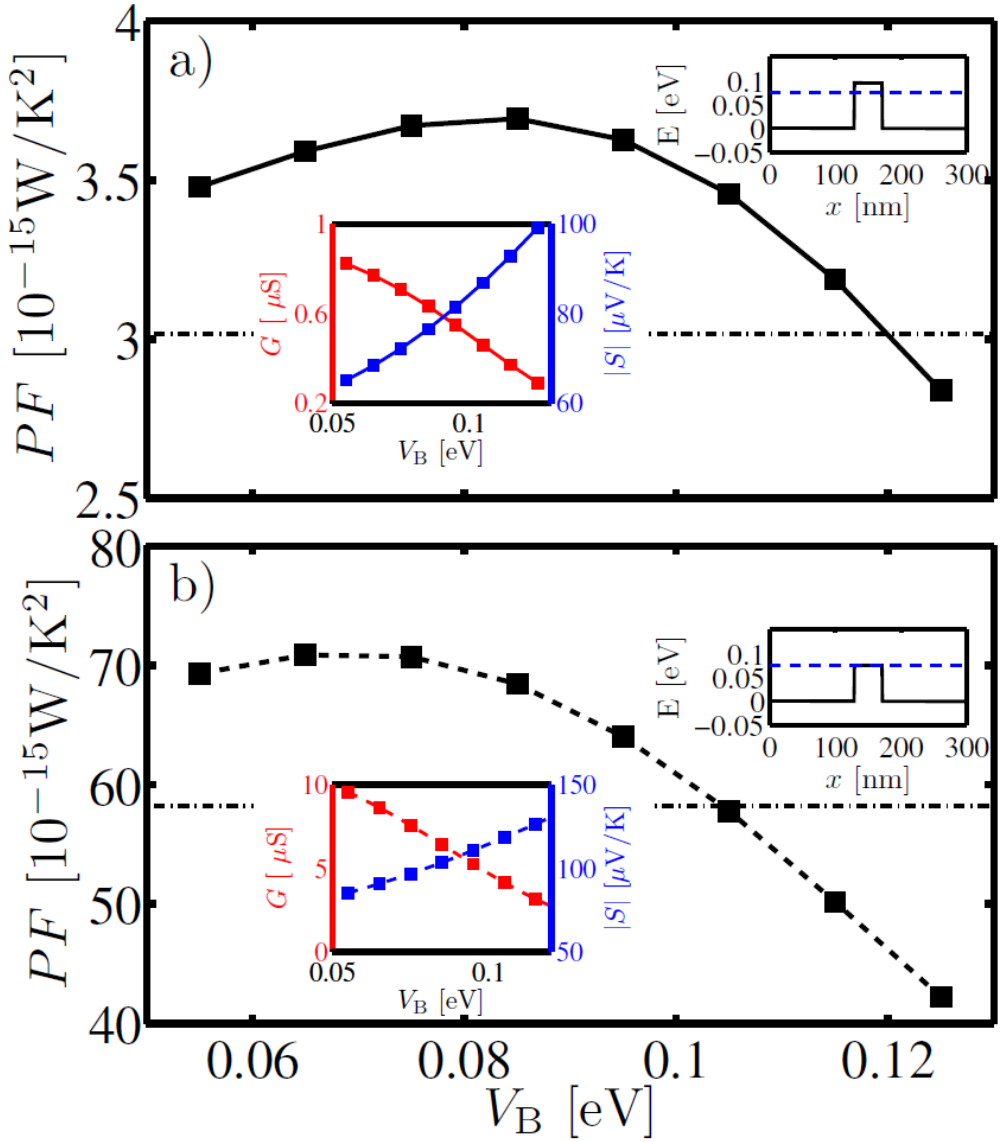

Figure 2 caption:

The TE power factor in (a) 1D, and (b) 2D of a channel of length $L = 300$ nm, with a single barrier of length $L_B = 42$ nm placed in the middle, versus the barrier height $V_B$. The Fermi level is at $E_F = 0.075$ eV. The dotted lines show the power factor of the empty channel (no barrier) for comparison. The insets show the electrical conductance and Seebeck coefficient versus $V_B$. The top-right schematics indicate the $V_B$ which maximizes the power factor in each channel. For 1D, $V_B \sim E_F + 10$ meV, and for 2D, $V_B \sim E_F - 10$ meV.

**Figure 3:**

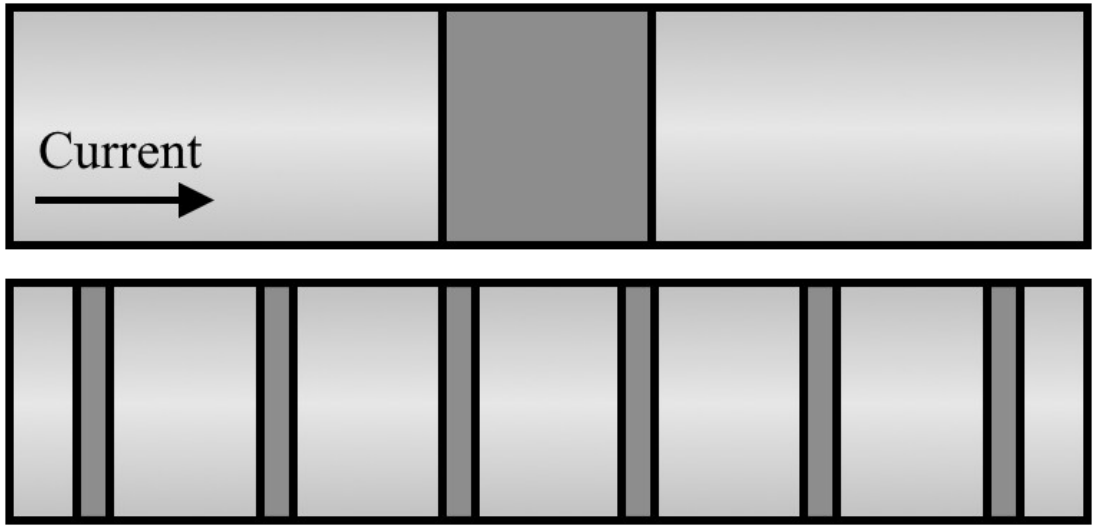

**Figure 3 Caption:**

Schematic diagrams of a superlattice channel with dark grey indicating a different barrier material. In (a) all of the material is concentrated in the center, and the thermoelectric properties are examined in Fig. 2. In (b) the same amount of barrier material is split into six smaller regions and spread over the channel, creating a series wells where carrier semi-relaxation can occur. The thermoelectric properties of this channel are described in Fig. 4.

Figure 4:

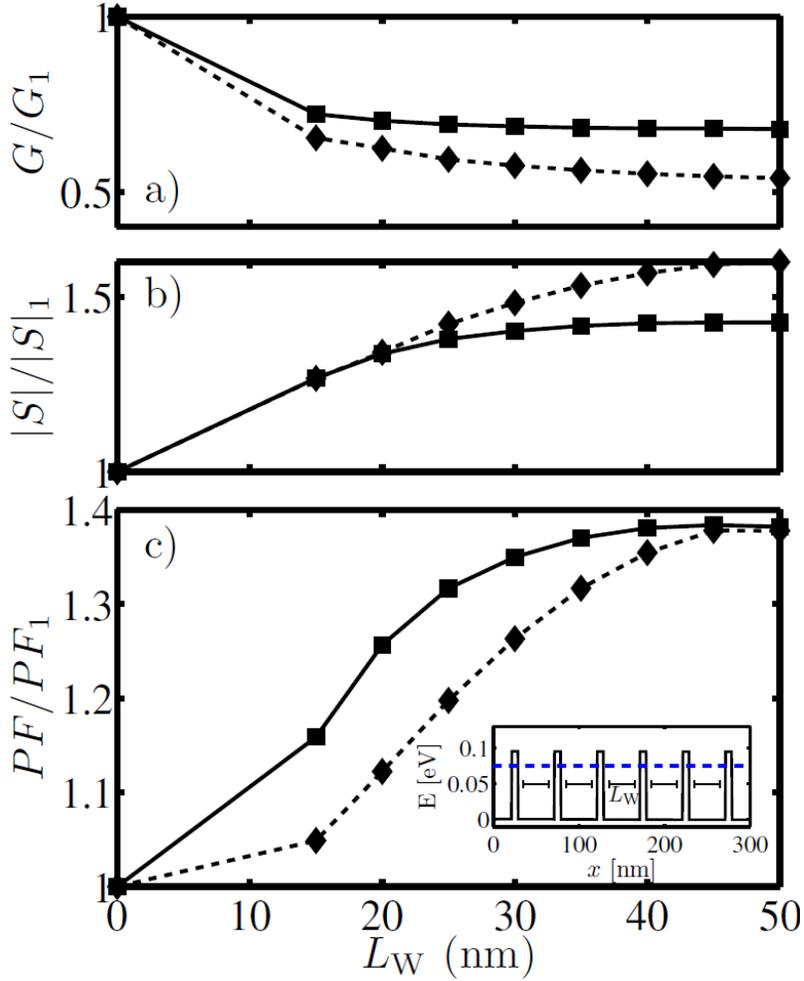

Figure 4 caption:

The *relative change* in the thermoelectric coefficients between a superlattice (SL) geometry composed of six barriers and the geometry where the entire 'barrier material' is centered in the middle as one wide barrier, versus the potential well length. A schematic of the SL geometry is shown in the inset of (c) which also indicates the meaning of $L_W$ as the spacing between barriers. (a) The change in the electrical conductance. (b) The change in the Seebeck coefficient. (c) The change in the power factor. The solid lines (squares) are results for 1D channels, whereas the dashed lines (diamonds) for 2D channels. The Fermi level is at $E_F = 0.075$ eV in all cases, and the $V_B$ is at the optimal PF conditions for the 1D SL ($V_B = E_F + k_BT$) and the 2D SL ($V_B = E_F$).

Figure 5:

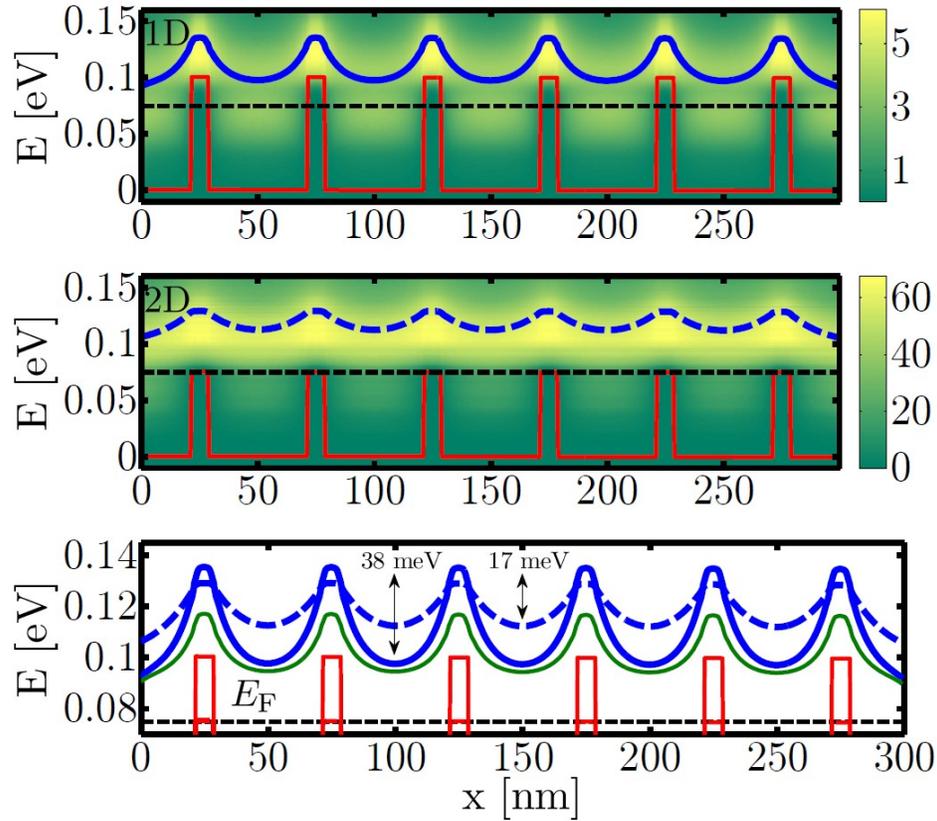

Figure 5 caption:

The current energy flow in a (a) 1D and (b) 2D SL of six barriers (red lines) and $L_W = 50$ nm. The Fermi level is at $E_F = 0.075$ eV in all cases, and the $V_B$ is at the optimal PF conditions for 1D ($V_B = E_F + k_BT$) and 2D ($V_B = E_F$). The blue lines show the average energy of the current flow. (c) The barriers and average energy (blue lines form (a) and (b)) are superimposed in order to provide a direct comparison. Under optimal PF conditions, the 1D channel has slightly higher current energy in the barrier regions, and lower current energy in the well regions compared to 2D, which shows much less variation (38 meV in 1D versus 17 meV in 2D). The green line shows the average current energy of the 1D channel with same barrier as the 2D channel ($V_B = E_F$), which indicates that for the same $E_F$ and $V_B$, the current energy in 1D is much lower compared to that in 2D.

Figure 6:

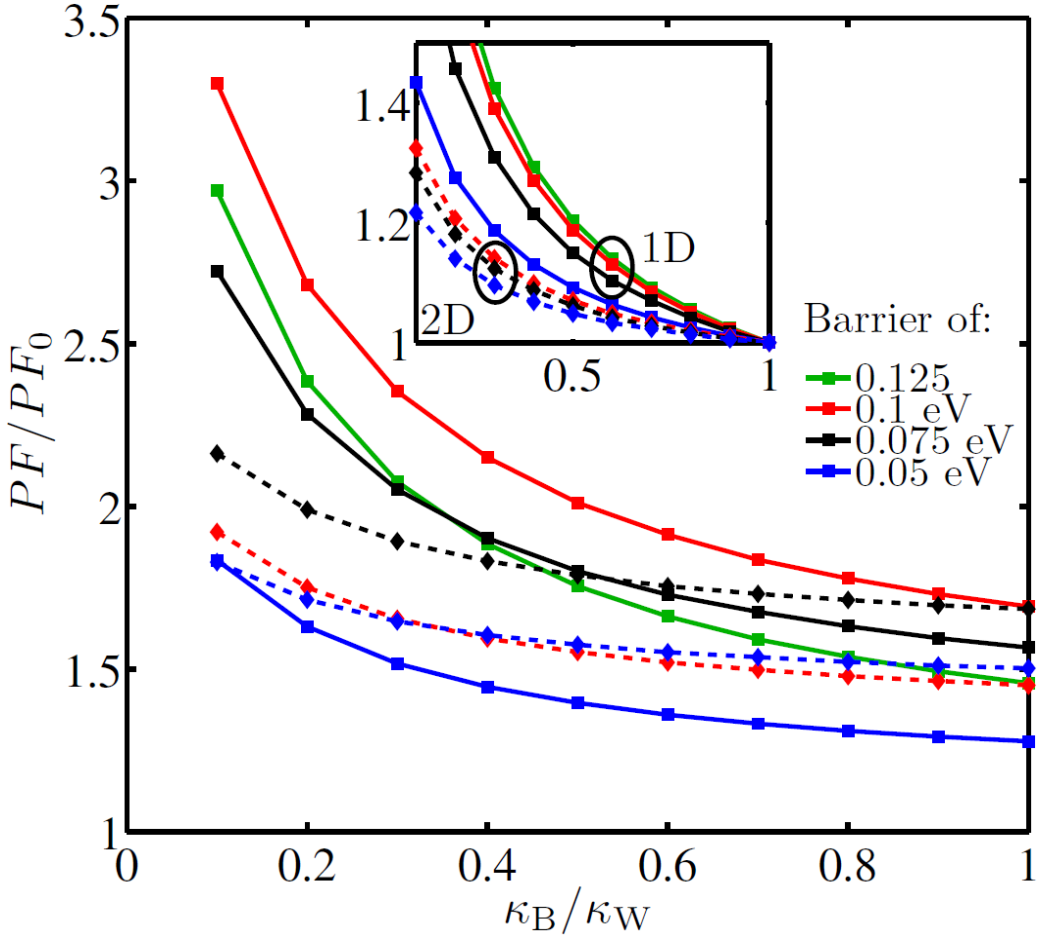

Figure 6 caption:

The relative change in the thermoelectric power factor between a superlattice which has different thermal conductivity in the barriers ($\kappa_B$) and wells ($\kappa_W$), compared to a superlattice with uniform thermal conductivity in all regions, versus the ratio of the thermal conductivities in the barriers and wells. The SL geometry considered has six barriers and $L_W$ = 50 nm. The solid lines are results for 1D channels, whereas the dashed lines for 2D channels. The Fermi level is at $E_F$ = 0.075 eV in all cases. Four different $V_B$ cases are considered: i) $V_B$ = 0.05 eV (blue lines, ~$k_B T$ lower than the 2D optimal), ii) $V_B$ = 0.075 eV (black lines, optimal PF conditions for 2D), $V_B$ = 0.1eV (red lines, optimal PF conditions for 1D), and $V_B$ = 0.125 eV (green line, ~$k_B T$ higher than the 1D optimal). Inset: The same data normalized to the $\kappa_B/\kappa_W$ = 1 value.

Figure 7:

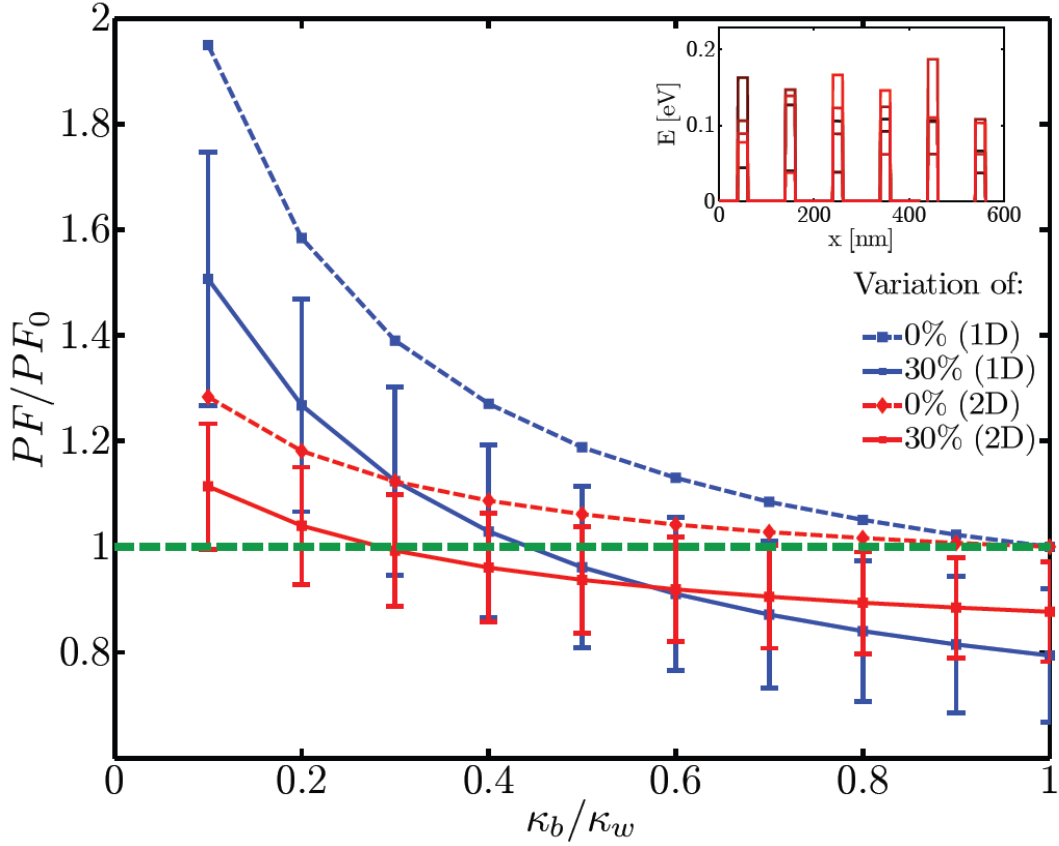

Figure 7 caption:

The relative change in the thermoelectric power factor between a superlattice with randomly varying barrier heights, which has different thermal conductivity in the barriers ($\kappa_B$) and wells ($\kappa_W$), normalized to a superlattice with uniform thermal conductivity in all regions $\kappa_B/\kappa_W = 1$, versus the ratio $\kappa_B/\kappa_W$. The SL geometry considered has six barriers and $L_W$ = 50 nm. The blue-dashed lines are results for 1D channels (with optimal $V_B$ = 0.1 eV), whereas the red-dashed lines for 2D channels (with optimal $V_B$ = 0.075 eV). The Fermi level is at $E_F$ = 0.075 eV in all cases. The solid lines with error bars indicate the power factor upon 30% variation in the barrier heights $V_B$ (blue-solid for 1D, and red-solid for 2D). In other words, barrier heights are drawn from a Gaussian distribution with a standard deviation which is 30% of the barrier height. 20 samples are used for each data point. Inset: Overalpping a few schematics of the SL geometry upon $V_B$ variations.